\documentclass[aip,amsmath,amssymb,12pt
]{revtex4-1}

\usepackage{amsfonts}
\usepackage{amsmath}
\usepackage{amssymb}
\usepackage{amsthm}
\usepackage{appendix}
\usepackage{datetime}
\usepackage{dcolumn} 
\usepackage{enumerate}
\usepackage[T1]{fontenc}
\usepackage{epigraph}
\usepackage[T1]{fontenc}
\usepackage{graphicx} 
\usepackage[utf8]{inputenc}
\usepackage{hyperref}
\hypersetup{
    colorlinks=true,
    linkcolor=blue,
    filecolor=magenta,      
    urlcolor=cyan,
}
\usepackage{mathptmx}
\usepackage{xcolor}

\newcommand{\R}{\mathbb{R}}  

\newcommand{\bfr}{{\bf r}}

\newcommand{\bfR}{{\bf R}}

\newcommand{\bra}[1]{\ensuremath{\langle #1 \vert}}
\newcommand{\ket}[1]{\ensuremath{\vert #1  \rangle}}

\newcommand{\op}[1]{\mathsf #1}

\DeclareMathOperator{\dd}{d}
\DeclareMathOperator{\erf}{erf}
\DeclareMathOperator{\erfc}{erfc}
\makeatletter
\newcommand{\thickbar}{\mathpalette\@thickbar}
\newcommand{\@thickbar}[2]{{#1\mkern1.5mu\vbox{
  \sbox\z@{$#1\mkern-1.5mu#2\mkern-1.5mu$}%
  \sbox\tw@{$#1\overline{#2}$}%
  \dimen@=\dimexpr\ht\tw@-\ht\z@-.8\p@\relax
  \hrule\@height.8\p@ 
  \vskip\dimen@
  \box\z@}\mkern1.5mu}
}
\makeatother
\renewcommand*{\bar}{\thickbar}
\newcommand{\andreas}[1]{{#1}}

\begin{document}

\title{Exploring the role of mean-field potentials and short-range wave function behavior in the adiabatic connection}

\author{Anthony Scemama}
\affiliation{Laboratoire de Chimie et Physique Quantiques (UMR 5626),
Université de Toulouse, CNRS, UPS, France}
\email{scemama@irsamc.ups-tlse.fr}

\author{Andreas Savin}
\affiliation{Laboratoire de Chimie Th\'eorique, CNRS and Sorbonne University \\ 4 place Jussieu, 75252 Paris, France}
\email{andreas.savin@lct.jussieu.fr}

\begin{flushright}
\begin{minipage}{0.5\textwidth}
\noindent
\begin{small}
 submitted for publication in: \textit{J. Comput. Chem.}
\end{small} 
\end{minipage} 
\end{flushright}

\begin{abstract}
In this article, we explore the construction of Hamiltonians with long-range
interactions and their corrections using the short-range behavior of the wave
function. A key aspect of our investigation is the examination of the
one-particle potential, kept constant in our previous work, and the
effects of its optimization on the adiabatic connection.

Our methodology involves the use of a parameter-dependent potential
dependent on a single parameter to
facilitate practical computations. We analyze the energy errors and densities
in a two-electron system (harmonium) under various conditions, employing
different confinement potentials and interaction parameters. The study reveals
that while the mean-field potential improves the expectation value of the
physical Hamiltonian, it does not necessarily improve the energy of the system
within the bounds of chemical accuracy.

We also delve into the impact of density variations in adiabatic connections,
challenging the common assumption that a mean field improves results.
Our findings indicate that as long as energy errors remain within chemical
accuracy, the mean field does not significantly outperform a bare potential.
This observation is attributed to the effectiveness of corrections based on the
short-range behavior of the wave function, a universal characteristic that
diminishes the distinction between using a mean field or not.
\end{abstract}


\maketitle

\tableofcontents
\newpage

\centerline{\today \hspace{2pt} at \currenttime}

\section{Motivation}

In a series of articles (the most recent being reference~\onlinecite{SavKar-JCP-23}), we utilized the short-range behavior of the wave function along an adiabatic connection to construct corrections to the energies obtained with Hamiltonians featuring long-range interactions.
This approach is dedicated to the application of range-separation methods where the short-range contribution of the electron-electron interaction is approximated.
The method does not involve any empirical or fitted parameters and is applicable to both ground and excited states.
However, the correction fails when the interaction in the Hamiltonian operates only at very large distances.

Until now, the one-particle local (external) potential was kept unchanged, diverging from the
common practice of initiating corrections from mean-field potentials. 
In this paper, we explore whether optimizing the potential could be beneficial in improving our approach.
A change in the potential can introduce a new term in the adiabatic connection.
Although we have exact expressions for short-range interactions, the need for such corrections raises the question of how to handle the new term in the adiabatic connection.

Another motivation for this study is density functional theory (DFT), based upon the theorem of Hohenberg and Kohn~\cite{HohKoh-PR-64}.
While the construction of an accurate universal density functional, applicable to all electronic systems, is feasible (as shown, {\em e.g.}, in ref.~\onlinecite{ColSav-JCP-99}), its practical realization is exceedingly challenging. 
The Hohenberg-Kohn theorem, however, does not provide a methodology for developing simple approximations.
Commonly, approximations are based on some assumptions about the density functional, which, owing to the universality of the exact functional, are then applied to various systems.
The adiabatic connection is a critical component in the construction of density functionals (see, {\em e.g.}, \onlinecite{StoGolPre-TCA-80,Bec-IJQC-83, ErnPer-JCP-98}, typically involving changes only in the two-particle density~\cite{LanPer-SSC-75}.

Recently, it has been repeatedly highlighted that while density functional approximations can yield reasonably accurate energies, they can result in densities of poor quality.~\cite{MarMedGerPanPerLys-MC-18, SimSonBur-JPCL-18, VucSonKozSimBur-19, SimSonSuhVucBur-22},
\andreas{In this paper, in contrast to density functional theory, we are not interested in obtaining the correct density from model calculations, but only the energy (for the ground and excited states).
The method we use can be extended to obtain expectation values of one-particle operators~\cite{Sav-JCP-20}, by using the Hellmann-Feynman theorem.
We do not deal with this subject in the present paper.
In fact, the density of our models are far from the exact density.}

It is also noteworthy that the hypothetical case of having an exact density for the Kohn-Sham system does ensure density invariance along the adiabatic connection.~\cite{HarJon-JPF-74}.
If the adiabatic connection includes a new term that is density-dependent, it necessitates the development of new approximations.

The results presented in this paper were obtained with Mathematica.\cite{Wol-23}

\section{Method}

\subsection{Schr\"odinger equation}
In this paper, we will consider Hamiltonians characterized by the following form
\begin{equation}
    \label{eq:h-model}
    H(v,\lambda,\mu) = T + V(\lambda,\mu) + W(\mu) + \lambda \bar{W}(\mu).
\end{equation}
Here, $T$ represents the kinetic energy operator. The term $V(\lambda,\mu) = \sum_{i=1}^{N} v(\bfr_i;\lambda,\mu)$ denotes a local one-particle potential, which notably depends on the parameters $\lambda$ and $\mu$. The symbol $N$ refers to the number of electrons within the system. The specific formulations for $v(\mathbf{r},\lambda,\mu)$ will be detailed subsequently. At this stage, it is essential to understand that when $\lambda=1$, $v(\mathbf{r},\lambda,\mu)$ is the external potential of interest (the ``physical'' one).

The electron-electron interaction is divided into two distinct components: $W(\mu)$ and $\bar{W}(\mu)$. These are defined as follows:
\begin{align}
\label{eq:w}
W(\mu) & = \sum_{i=2}^N \sum_{j=1}^{i-1} w(r_{ij}, \mu), \\
\label{eq:wbar}
\bar{W}(\mu) & = \sum_{i=2}^N \sum_{j=1}^{i-1} \bar{w}(r_{ij}, \mu),
\end{align}
where the individual interaction terms are given by
\begin{align}
w(r,\mu) & = \frac{\erf(\mu r)}{r}, \\
\bar{w}(r,\mu) & = \frac{\erfc(\mu r)}{r}.
\end{align}
In these equations, $r_{ij}$ represents the distance between electrons $i$ and $j$, and $\mu$ is a parameter that characterizes the interaction.

It is important to note that for any finite value of $\mu$, $W(\mu)$ does not represent the physical (Coulomb) interaction potential.
Instead, it is a model interaction, defined by the parameter $\mu$.
However, the sum of $w(r,\mu)$ and $\bar{w}(r,\mu)$, which equals $1/r$, corresponds to the Coulomb interaction.
Here, $w$ represents the long-range part of the interaction, while $\bar{w}$ accounts for the short-range component.
Consequently, when $\lambda=1$, the potential $v(\mathbf{r}, \lambda, \mu)$ becomes $v_1(\mathbf{r})$, and the Hamiltonian $H(v,\lambda=1,\mu)$ becomes $H_1(v_1)$, independent of the value of $\mu$. The range of the interaction is characterized by $1/\mu$.

The process of turning on the parameter $\lambda$ from 0 to 1 serves to transform the model into the physical system.
In this context, $\lambda$ functions as an adiabatic connection constant.

The Schr\"odinger equation is
\begin{equation}
 \label{eq:se}
  H(v,\lambda,\mu) \Psi(v,\lambda,\mu) = E(v,\lambda,\mu) \Psi(v,\lambda,\mu)
\end{equation}
When $\lambda=1$, we have $E_1(v_1)=E(v_1,\lambda=1,\mu)$, independently of $\mu$.

\subsection{Adiabatic connection}
We do not attribute any physical significance to $E(v,\lambda,\mu)$ except when $\lambda=1$.
However, we consider that the Schr\"odinger equation for $\lambda=0$ and a chosen model ($\mu$) is accessible.
They are related through the adiabatic connection formula,
\begin{equation}
    \label{eq:ac}
    E_1(v_1) = E(v,\lambda=0,\mu) + \int_0^1 \dd \lambda \, \partial_\lambda E(v, \lambda, \mu)
\end{equation}
for any $\mu \ge 0$.
The advantage of using the adiabatic connection appears when applying the Hellmann-Feynman theorem.
Recalling that $v$ is a function of $\lambda$ and $\mu$,
\begin{equation}
\label{eq:helfey}
\partial_\lambda E(v, \lambda, \mu)  =  \bra{\Psi(v,\lambda,\mu)} \partial_\lambda V(\lambda,\mu) \ket{\Psi(v,\lambda,\mu)} + \bra{\Psi(v,\lambda,\mu)} \bar{W}(\mu) \ket{\Psi(v,\lambda,\mu)}
\end{equation}
We rewrite eq.~\eqref{eq:ac} using $H_1$,
\begin{align}
\label{eq:ac-h}
\nonumber
E_1 = & \bra{\Psi(v,\lambda=0,\mu)} H_1(v_1) \ket{\Psi(v,\lambda=0,\mu)} \\ \nonumber
      &+\int_0^1 \dd \lambda \left[
       \bra{\Psi(v,\lambda,\mu)} \partial_\lambda V(\lambda,\mu) \ket{\Psi(v,\lambda,\mu)} - \bra{\Psi(v,\lambda=0,\mu)} \partial_\lambda V(\lambda,\mu) \ket{\Psi(v,\lambda=0,\mu)}
       \right] \\
      &+\int_0^1 \dd \lambda \left[  \bra{\Psi(v,\lambda,\mu)} \bar{W}(\mu) \ket{\Psi(v,\lambda,\mu)} - \bra{\Psi(v,\lambda=0,\mu)} \bar{W}(\mu) \ket{\Psi(v,\lambda=0,\mu)} \right]
\end{align}
The first term on the r.h.s is accessible through eq.~\eqref{eq:se}, the second term depends on the evolution of the local potential, and the third term, on that of the interaction.

Apparently, eq.~\eqref{eq:helfey} is useless, because we do not know $\Psi(v,\lambda,\mu)$ except for $\lambda=0$.
It was shown ({\em e.g.}, in refs. \onlinecite{Sav-JCP-11} or \onlinecite{SavKar-JCP-23}) that, for large $\mu$, the last term on the r.h.s. can be easily approximated by expressions using the behavior of the wave function for small $|\bfr_1 - \bfr_2|$ and the expectation value of $\bar{W}$ with $\Psi(v,\lambda=0,\mu)$.
We still have to deal with the second term on the r.h.s. of eq.~\eqref{eq:helfey}.

\subsection{The evolution of the local potential in the adiabatic connection}

In our previous works, we made the choice to keep $v$ independent of $\lambda$ and $\mu$. Specifically, we set $v(\bfr,\lambda,\mu) = v_1(\bfr_1)$ which is the ``physical'' or ``bare'' potential. This choice ensures that we recover the correct result when $\lambda=1$:
\begin{equation}
 \label{eq:v-bare}
 v_{\text{bare}}(\bfr) = v_1(\bfr).
\end{equation}

The condition $\partial_\lambda v(\mathbf{r},\lambda,\mu)=0$ leads to the elimination of the second term on the right-hand side of eq.~\eqref{eq:ac-h}.
Consequently, only the electron-electron interaction term varies along the adiabatic connection.


The choice to keep the potential constant with respect to $\lambda$ and $\mu$ may seem unorthodox at first glance, especially considering that a system characterized by a weak interaction could significantly diverge from the actual physical system.
However, this approach is balanced by applying a specific correction. The correction employed for the short-range electron-electron interaction is based on the short-range behavior of the wave function. This aspect is universal, implying that the influence of the external potential is negligible in this context.

Formally, there are alternative ways to eliminate the term dependent on the evolution of the local potential.
In DFT, the adiabatic connection typically maintains a constant density $\rho(\mathbf{r}, v, \lambda, \mu)$, generated by $\Psi(v,\lambda,\mu)$, which does not vary with $\lambda$ (and thus is also independent of $\mu$): $\rho(\mathbf{r}, v, \lambda,\mu) = \rho(\mathbf{r}, v_1, \lambda=1,\mu) \equiv \rho_1(\mathbf{r},v_1)$.
This necessitates an appropriate choice of $v$,\andreas{\cite{Lie-IJQC-83}}
\begin{equation}
 \label{eq:v-dft}
 v_{\text{DFT}}(\lambda,\mu) = \arg \max_{\tilde{v}} \left[ E(\tilde{v}, \lambda, \mu) - \int_{\R^3} \dd \bfr \,  \rho_1(\bfr,v_1) \tilde{v}(\bfr) \right] ,
\end{equation}

Such an approach was recently used for harmonium, the system studied in this paper, in ref.~\citenum{Hou-24}.
Although this approach is, in principle, exact, we do not adopt this DFT-inspired approach in the present paper for two reasons.
Firstly, it demands significant computational effort.
Secondly, it is not necessary \andreas{as we are only interested in obtaining the energy, not the density.}
Indeed, if $v$ depends on a parameter $\omega$, we can limit the search to
\andreas{
\begin{equation}
 \label{eq:omega-opt}
 \omega_{\text{opt}}(\lambda,\mu) = \arg \max_{\omega,} \left[ E(v(\omega), \lambda, \mu) - \int_{\R^3} \dd \bfr \,  \rho_1\left(\bfr, v(\omega_1)\right) v(\bfr, \omega, \lambda, \mu) \right].
\end{equation}
}
Here, $\omega_1$ is the value of $\omega$ that recovers $v_1$.
The optimal potential $v_{\text{opt}}(\lambda,\mu)$ is obtained with $\omega_{\text{opt}}$.
Applying the Hellmann-Feynman theorem, we obtain
\begin{equation}
 \label{eq:v-dft-helfey}
 \left(
  \int_{\R^3} \dd \bfr \,  \rho(\bfr,v,\lambda,\mu) \partial_\omega v(\omega, \bfr, \lambda, \mu) -  \int_{\R^3} \dd \bfr \,  \rho_1(\bfr) \partial_\omega v(\omega, \bfr, \lambda, \mu) \right)_{\omega=\omega_{\text{opt}}} = 0
\end{equation}
For the adiabatic connection, the parameter $\omega$ depends on $\lambda$.
Multiplying eq.~\eqref{eq:v-dft-helfey} by the derivative of $\omega$ with respect to $\lambda$, we find that the integrand of the second term on the right-hand side of eq.~\eqref{eq:ac-h} vanishes for all $\lambda$.
Therefore, it is not necessary to produce $\rho$ for $0 < \lambda < 1$.
However, knowledge of $\rho$ at $\lambda=1$ is still required, as it determines, through eq.~\eqref{eq:omega-opt}, $v_{\text{opt}}(\lambda=0,\mu)$, i.e., the model system.
Also with this choice, only the interaction term varies along the adiabatic connection.

In order to deal with a practically accessible potential, we consider, as above, a potential that depends on a parameter, $\omega$.
However, the determination of $\omega$ is now guided by the following criterion:
\begin{equation}
 \label{eq:omega-oep}
 \omega_{\text{OEP}}(\lambda=0,\mu) = \arg\min_\omega \bra{\Psi\left( v(\omega), \lambda=0, \mu \right)} H_1 \ket{\Psi\left( v(\omega), \lambda=0, \mu \right)} .
\end{equation}
This is analogous to the definition of the optimized effective potential~\cite{AasLukTal-ADNDT-78}.
 For values of $\lambda$ greater than zero, we employ a linear interpolation of the potentials:~\cite{HarJon-JPF-74}:
\begin{equation}
 \label{eq:v-oep}
 v_{\text{OEP}}(\bfr_,\lambda,\mu)=(1-\lambda)v_{\text{OEP}}\left(\bfr_,\omega_{\text{OEP}}(\lambda=0,\mu)\right)+ \lambda v_1(\bfr)
\end{equation}

In this framework, the evolution of the potential along the adiabatic connection contributes to the potential-dependent term in eq.~\eqref{eq:ac-h}.

An illustrative example of this approach is provided in Appendix~\ref{app:non-int}.
In this example, both $\omega_{\text{opt}}$ and $\omega_{\text{OEP}}$ are calculated for a system of two electrons in a harmonic confinement at $\lambda=\mu=0$.

\subsection{The evolution of the interaction in the adiabatic connection}

In addressing the variation of the interaction within the adiabatic connection, specifically the last term on the right-hand side of eq.~\eqref{eq:ac-h}, we adopt the procedure outlined in our previous publications.~\cite{Sav-JCP-20} 
This approach is based on the observation that for large values of $\mu$, the short-range behavior of the wave function becomes predominantly significant.
To derive corrections to the model, we focus on the behavior of the wave function at short range (as $r \rightarrow 0$) in the context of large $\mu$ values, as detailed in ref.~\onlinecite{GorSav-PRA-06}.
This consideration leads us to the following expression:
\begin{align}
 \nonumber
 & \int_0^1 \dd \lambda \left[  \bra{\Psi(v,\lambda,\mu)} \bar{W}(\mu) \ket{\Psi(v,\lambda,\mu)} - \bra{\Psi(v,\lambda=0,\mu)} \bar{W}(\mu) \ket{\Psi(v,\lambda=0,\mu)} \right]  \\
 & = \alpha(\mu) \bra{\Psi(v,\lambda=0,\mu)} \bar{W}(\mu) \ket{\Psi(v,\lambda=0,\mu)}
  \label{eq:ac-wbar}
\end{align}
where $\alpha$ is a simple function of $\mu$, 
and its specific form is provided in ref.~\onlinecite{SavKar-JCP-23}.

\subsection{The system studied}

To illustrate our approach, we use data from a two-electron system known as harmonium.
The rationale behind choosing this system lies in the ease of obtaining accurate data, which in turn allows for a precise evaluation of the errors introduced by various approximations.
In the case of harmonium, the external potential is defined as:
\begin{equation}
 \label{eq:v-harm}
 v_1(\bfr) = \frac{1}{2} \omega_1^2 r^2.
\end{equation}
The corresponding electron density for this system is either analytically known (for $\omega_1=1/2$)~\cite{KaiHerHanMur-93} or can be numerically generated.
For cases where $\lambda \neq 1$, we adopt the potential $v(\mathbf{r},\lambda,\mu) = \frac{1}{2} \omega(\lambda,\mu)^2 r^2$ and determine $\omega$ using the three variants previously discussed: $\omega_{\text{bare}}$, $\omega_{\text{opt}}$, and $\omega_{\text{OEP}}$.

While this method is applicable to excited states as well (as discussed in, for example, ref.~\onlinecite{Sav-JCP-20}), in this paper, we focus on the ground state of the harmonium system.
\andreas{We give results for an excited state of harmonium only in Appendix~\ref{app:excited}.
}

\andreas{
Harmonium has special features, that are not present in atomic or molecular systems.
Electrons cannot escape its infinite confinement.
The energy level ordering is different from that in atoms, and their spacing is different.
However, we are interested in (a part of) the correlation energy, and this is very close in harmonium to that in the He series.
Therefore, we do not expect the difference in the ordering of the energy levels to have an impact on our conclusions.
}

We cannot expect eq.~\eqref{eq:ac-wbar} to be effective for small values of $\mu$ because the prefactor $\alpha(\mu)$ is determined for $\mu \rightarrow \infty$.
Specifically, when $\mu=0$, the energy of the model system, $E(v,\lambda,\mu)$, deviates significantly from the physical energy, which is defined by $\omega_1$.
For instance, with $\omega_1=1/2$, the energy difference is 0.5 hartree below the exact result of 2 hartree.
It also happens to be the error of the exact Kohn-Sham system.

As $\mu$ increases, the energy of the model system gradually converges towards the exact energy.
We know that for large $\mu$, the error in the energy decreases at a rate proportional to $\mu^{-4}$.~\onlinecite{Sav-JCP-20}

\begin{figure}
    \centering
    \includegraphics{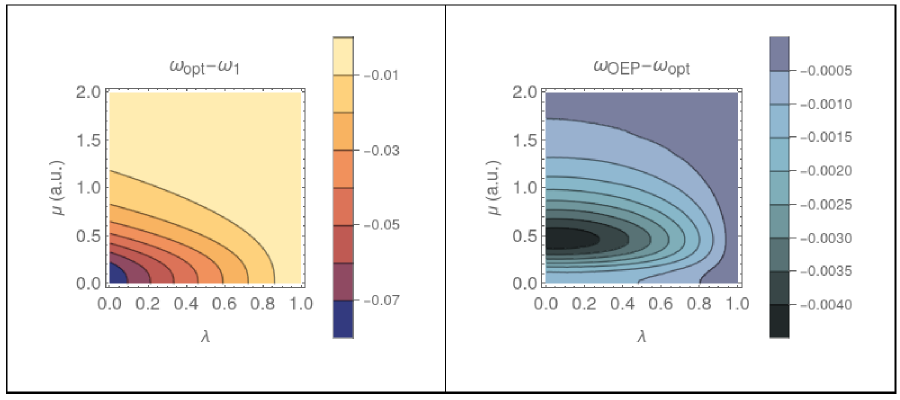}
    \caption{The difference between the optimized potential parameter $\omega_{\text{opt}}$ and $\omega_1=0.5$ (a.u., left panel) and between $\omega_{\text{OEP}}$ and $\omega_{\text{opt}}$ (a.u., right panel); see eqs~\eqref{eq:omega-opt}, \eqref{eq:omega-oep}, and \eqref{eq:v-oep}.}
    \label{fig:omegas}
\end{figure}
Fig.~\ref{fig:omegas} illustrates in the left panel that the confining parameter $\omega_{\text{opt}}$, as defined in eq.~\eqref{eq:omega-opt}, tends to decrease in systems with weaker repulsion. 
This trend is logical, considering that the aim here is to approximate the exact density, which becomes more dispersed due to repulsion.
The most pronounced effect is observed in the system with no interaction ($\lambda=0$, $\mu=0$).

In the right panel of fig.~\ref{fig:omegas}, we observe that the confining parameter determined through the OEP scheme, as per eqs.~\eqref{eq:omega-oep} and \eqref{eq:v-oep}, aligns closely with the parameter that cancels out the one-body term in the adiabatic connection.

\section{Results}

\subsection{Densities}
\begin{figure}
    \centering
    \includegraphics{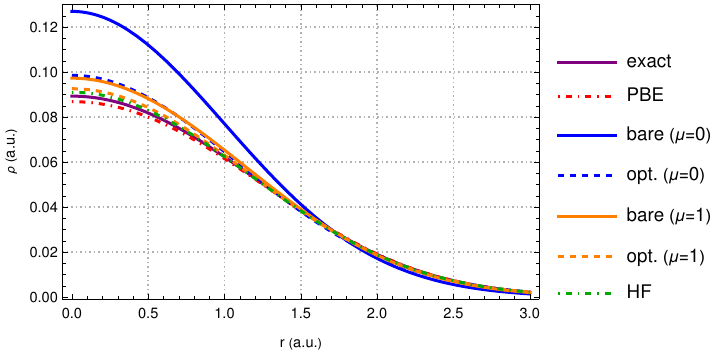}
    \caption{Densities obtained for the non-interacting system ($\lambda=0, \mu=0$) with the bare potential, $\omega_{\text{bare}}=\omega_1=1/2$, represented by the full blue curve, and the optimized potential, $\omega_{\text{opt}}$ as per eq.~\eqref{eq:omega-opt}, shown with a full orange  curve. The densities obtained with these potentials for the model systems with $\mu=1$ are shown as dashed curves. These are compared to the exact density, depicted by the full purple curve, \andreas{and the Hartree-Fock and PBE densities, illustrated with dot-dashed green and red curves.}}
    \label{fig:densities}
\end{figure}
We first would like to point out that the densities in our description can experience significant changes, as illustrated in fig.~\ref{fig:densities}.
For comparison, we have juxtaposed the exact density with the density produced by \andreas{the Hartree-Fock and Perdew-Burke-Ernzerhof (PBE)\cite{PerBurErn-PRL-96} approximations.}
Notably, these two densities exhibit a close resemblance.
\andreas{
Thus, the density effect we are observing here is  stronger than that used in density-corrected DFT~\cite{SimSonBur-JPCL-18,VucSonKozSimBur-19,SimSonSuhVucBur-22}}. 

However, completely neglecting the electron-electron interaction (by using $\omega_{\text{bare}}$ at $\lambda=0, \mu=0$) leads to a substantial deviation.
The optimization of $\omega$ partially rectifies this discrepancy, but since the potential has not been fully optimized, as indicated by the comparison between eqs.~\eqref{eq:v-dft} and \eqref{eq:omega-opt}, the exact density is not recovered.
Despite these significant effects on the density, it is interesting to note that the second term on the right-hand side of eq.~\eqref{eq:ac-h} still vanishes.
The density obtained using $\omega_{\text{OEP}}$ is not depicted in the figure, as it is nearly indistinguishable from that obtained with $\omega_{\text{opt}}$ on the scale of the plot.
We also see in fig.~\ref{fig:densities} that while approaching the exact system, {\em e.g.}, at $\mu=1$ the densities can still show a noticeable difference from the true one.

\subsection{Energy errors}
\begin{figure}
    \centering
    \includegraphics[width=0.6\textwidth]{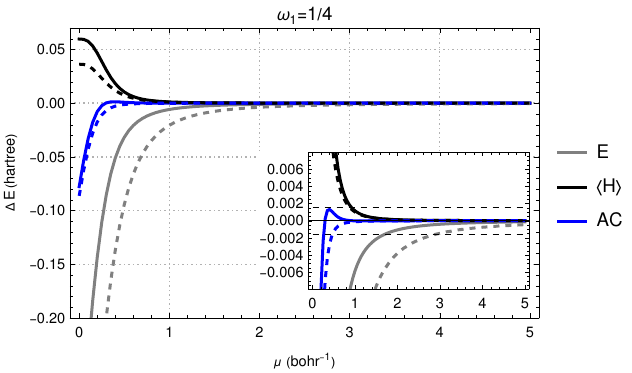} \\
    \includegraphics[width=0.6\textwidth]{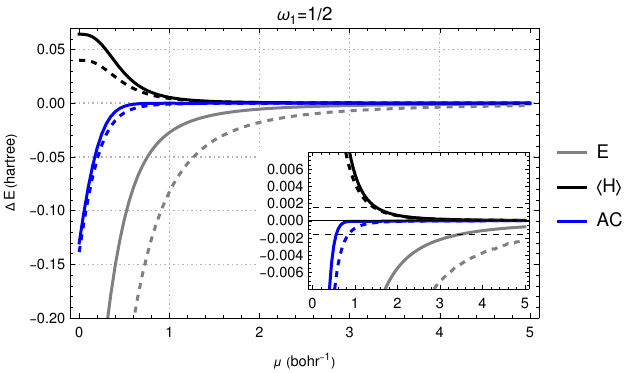} \\
    \includegraphics[width=0.6\textwidth]{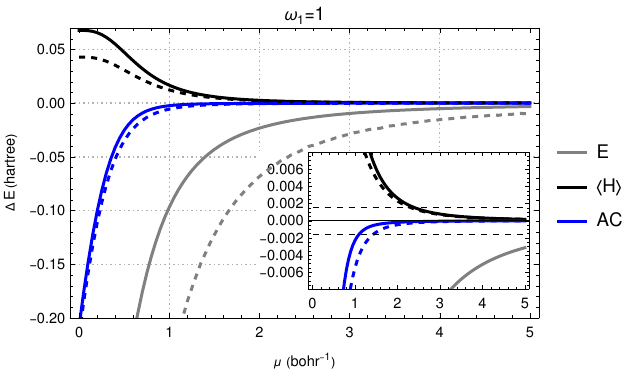}
    \caption{Energy errors for a quadratic confinement potential, as defined in eq.~\eqref{eq:v-harm}, with $\omega_1=$ 1/4 (top panel), 1/2 (middle panel), and 1 (bottom panel), plotted as functions of the parameter defining the interaction in the model, $\mu$, as per eq.~\eqref{eq:w}.
Full curves represent results obtained with the bare potential, $\omega_{\text{bare}}$, and dashed curves for those obtained with $\omega_{\text{OEP}}$.
The model energies are shown in gray, the expectation values of the Hamiltonian with the physical potential ($v_1, 1/r$) in black, and those after correction using the adiabatic connection and the asymptotic behavior of the wave function, as per eq.~\eqref{eq:ac-wbar}, in blue.
The inset displays the same curves over a narrower energy range.
Horizontal dashed lines indicate the range of chemical accuracy (errors within 1kcal/mol).}
    \label{fig:energies}
\end{figure}

In this section, we examine the impact of employing a ``mean-field'' potential.
We consider three different settings for $v_1$ in eq.~\eqref{eq:v-harm}, choosing $\omega_1=$ 1/4, 1/2, and 1.
For the one-particle potential $v(\bfr,\omega,\lambda,\mu)$, fig.~\ref{fig:energies} presents results obtained using the bare potential (full curves) and those obtained with $v_{\text{OEP}}$.
Those obtained with $v_{\text{opt}}$ are not shown in this figure, as they are very close to those obtained with $v_{\text{OEP}}$.
The largest difference observed was for $\omega_1=1$ and $\mu \approx 0.5$ and is $\approx 0.2$~mhartree.

Three approximations are employed to calculate the energy.
Firstly, we simply use $H(v,\lambda=0,\mu)$ as defined in eq.~\eqref{eq:h-model}.
Secondly, we compute the expectation value of the physical Hamiltonian, $H(v_1)$, using $\Psi(v,\lambda=0,\mu)$ (see eq.~\eqref{eq:se}).
Thirdly, we apply the adiabatic correction that becomes exact for large $\mu$, as per eq.~\eqref{eq:ac-wbar}.
The corresponding energy errors are depicted in fig.~\ref{fig:energies} as gray, black, and blue curves, respectively.

Figure~\ref{fig:energies} illustrates how the energy errors evolve as the interaction in the model is progressively increased (i.e., as $\mu$ increases).
Generally, we anticipate that approximations will be more cost-effective for smaller values of $\mu$.
Therefore, our focus is on achieving accurate results at lower $\mu$ values, balancing computational efficiency with the precision of the model.

The first observation from our analysis is that the model energies exhibit substantial errors.
Replacing $v_1$ with a mean field potential ($v_{\text{OEP}}$) does not lead to an improvement in these energies.
It's important to note that the purpose of the mean field is to improve the expectation value of the physical Hamiltonian, rather than directly improving the energy of the model.

Indeed, the expectation value of the physical Hamiltonian is consistently improved by the mean field across all cases.
This results are in error of about 0.04 hartree.
For a two-electron system, this is approximately the magnitude of the correlation energy.

As expected, introducing some interaction into the model reduces this error.
Furthermore, applying the adiabatic correction, which is valid for large values of $\mu$, naturally improves the results for these larger values.
In our examples, $\mu$ can be considered ``large enough'' already if greater than 1 bohr$^{-1}$. 
\andreas{As $\mu$ is an inverse distance, the separation between short and long range occurs at an inter-electronic distance of $1/\mu$.
}
Focusing on the region of chemical accuracy (errors within $\pm 1$ kcal/mol) as defined in~\cite{Pop-RMP-99}, we observe that the adiabatic connection performs exceptionally well for large $\mu$, yielding relatively flat curves.
However, as $\mu$ decreases, there is a noticeable decline in accuracy.
In every case we examined, applying the adiabatic connection reduces the value of $\mu$ for which the absolute error exceeds 1 kcal/mol.
Yet, within the bounds of chemical accuracy, the mean field does not demonstrate any significant beneficial effect.

\andreas{One can notice a small peak in the error of our approximation around $\mu \approx 0.4$~bohr$^{-1}$ for $\omega=1/4$~a.u.
Such a bump occurs also for the first excited state at $\omega=0.5$~a.u.
We do not know how to predict when such a bump occurs (see app.~\ref{app:excited}).
We believe it is related to our asymptotic treatment of the wave function.
It is related to an expansion in powers of the inter-electronic distance that are not treated correctly in the approximation presented in this paper.
When higher powers of the inter-electronic distance, $r$, are taken into account, the sign of the error changes when $\mu$ (as a measure of $1/r)$  becomes small (see fig. 3 in ref.~\citenum{SavKar-JPCA-23}).~\footnote{As an analogy, consider cutting off the Taylor expansion of $\exp(-x)$ at small powers of $x$.}
}

\section{Conclusion}

Intuitively, one might anticipate that changes in density would play a crucial role in the adiabatic connection process.
This expectation aligns with the prevalent understanding that employing a mean field for the zeroth-order Hamiltonian enhances the quality of the results.
We found that as long as the energy errors remained within the bounds of chemical accuracy, the use of a mean field did not demonstrate any advantage over a bare field.
A plausible explanation for this observation is that the most accurate results in this region were achieved using a correction that accounts for the short-range behavior of the wave function.
Since this behavior is universal, the choice between using a mean field or not does not significantly alter the results.
These results do not support the necessity of applying a one-body correction to the adiabatic connection \andreas{for obtaining the energy}, even when the density significantly deviates from the exact one.
This finding is somewhat disappointing, as it does not suggest any improvement by employing a mean-field potential. However, the study is numerical and focuses on a simple system, necessitating caution in generalizing these results.


\andreas{Our methodology differs from density functional approximations by the use of only exact, universal properties of the wave function.
The price to pay is that we have to keep a non-negligible interaction between electrons.
Of course, one can try to use our findings in constructing density functional approximations, or other approximations that take into account the behavior for weak interactions, such as what we have termed the second-order adiabatic correction~\cite{SavKar-JCP-23}.
These might demonstrate a more important impact of the mean field.
The exploration of these potential effects and their implications remains an open area for future research.}

\andreas{Another difference to density functional approximations is that we do not aim to obtain accurate densities. 
Our densities are far from the correct ones.
We claim not having observed an effect of the mean field on the energy approximations, as long as they are accurate, but this does not preclude the utilization of mean fields.
Furthermore, properties can be obtained not through expectation values, but as perturbations, and thus from energy expressions.
}

\andreas{A problem known already with the introduction of range-separation (see Ewald's paper~\cite{Ewa-AP-21}), is also present in density functional approximations~\cite{PolSavLeiSto-JCP-02}, and in our paper. 
We did not treat it in this paper, but it deserves to be mentioned.
It is the question of choosing the correct range-separation parameter, $\mu$, to find the balance the computational effort and the accuracy.
Like with basis sets, experience can decide about such a value.
Typically, it seems that our $\mu$ should lie between 0.5 and 1~bohr$^{-1}$.
However, the optimal value is system- and state-dependent.~\cite{KarSav-PTCP-23}
We see this also in fig.~\ref{fig:energies} that our approximation starts to work well (the electrons are close) at different critical values of $\mu \approx 0.3$, $0.6$ and $1.1$~bohr$^{-1}$, for increasing confinement ($\omega = 0.25$, $0.5$ and $1$~a.u., respectively).
This parallels the expectation values of $1/r_{12}$ that is, for the physical system, of $\approx 0.29$, $0.45$ and $0.67$~hartree.
However, choosing different values for different systems or states can produce problems with size-consistency.
A proposed solution was to define $\mu$ locally.
Efficient algorithms are already known~\cite{KlaBah-JCTC-20} and being applied
(see, e.g., ref.~\citenum{BruBahKum-JCP-22}).
It has been also noticed that basis sets provide, in fact, a local $\mu$.~\cite{GinPraFerAssSavTou-JCP-18,LooPraSceTouGin-JPCL-19,GinSceLooTou-JCP-20,TraGinTou-JCP-23}
}

\section{Acknowledgement}
\andreas{This paper is dedicated to Carlo Adamo.
His work is well-known for the many interesting applications of density functional theory.
However, he is interested in using rigorously defined approximations, including hybrids and double hybrids.
Our method is related to these methods, in the sense that it uses correlated wave functions, and discusses optimized effective potentials.}

\appendix

\section{$\omega_{opt}$ and $\omega_{\text{OEP}}$ in the non-interacting 
  ($\lambda=0, \mu=0$) limit of harmonium}
\label{app:non-int}

In the non-interacting limit, $\lambda=0$, $\mu=0$, for a harmonic confinement potential specified by $\omega$, we have
\begin{equation}
    \label{eq:Phi}
    \Phi(\omega, \bfR, \bfr)=\Psi(v(\omega),0,0)=(\omega/\pi)^{3/2} e^{-\omega (R^2+r^2/4)} .
\end{equation}
where $\bfR=(\bfr_1+\bfr_2)/2$ is the center-of-mass coordinate, and $\bfr=\bfr_1-\bfr_2$ is the relative coordinate; $R=|\bfR|$, $r=|\bfr|$.
The Hamiltonian of the non-interacting problem is:
\begin{equation}
    \label{eq:h0}
    H_0\left( v(\omega) \right) = -\frac{1}{4}\left( \partial_R^2 +\frac{2}{R} \partial_R \right) +\omega^2 R^2 -\left( \partial_r^2 +\frac{2}{r} \partial_r \right) +\frac{1}{4} \omega^2 r^2 .
\end{equation}
$E\left(v(\omega),0,0 \right)$, showing up in eq.~\eqref{eq:omega-opt} is $3 \omega$.
The physical density for $\omega_1=1/2$ is analytically known~\cite{KaiHerHanMur-93},
\begin{equation}
    \label{eq:rho-exact}
    \rho_1(\bfr)=\frac{2}{\pi (5 \pi + 8 \sqrt{\pi})}
      \left( e^{-r^2} +\frac{\sqrt{2 \pi}}{8}e^{-r^2/2} 
      \left( 7+r^2+ 4(1/r+r)\erf(r/\sqrt{2}) \right) \right) .
\end{equation}
The function to be maximized over $\omega$ in eq.~\eqref{eq:omega-opt} is
\begin{equation}
    \label{eq:get-opt}
    3 \omega - \frac{1}{2} \omega^2 \int_{\R^3} \dd \bfr \, \rho_1(\bfr) r^2 =  3 \omega + \frac{ 56+36 \sqrt{\pi} }{2 (8+5 \sqrt{\pi )} } \omega ^2 .
\end{equation}
It yields $\omega_{\text{opt}} \approx 0.4222$.

In order to obtain $\omega_{\text{OEP}}$, eq.~\eqref{eq:omega-oep}, we minimize over $\omega$
\begin{equation}
    \label{eq:get-oep}
    \bra{\Phi(\omega)} H_0 \left( v(\omega_1) \right) + 1/r \ket{\Phi(\omega)} =\frac{3}{2}\omega + \frac{3}{2}\frac{\omega_1^2}{\omega} + \sqrt{\frac{2}{\pi}} \sqrt{\omega}
\end{equation}
It yields $\omega_{\text{OEP}} \approx 0.4211$, very close to $\omega_{\text{opt}}$.

\section{An open shell example}
\label{app:excited}
\begin{figure}
    \centering
    \includegraphics{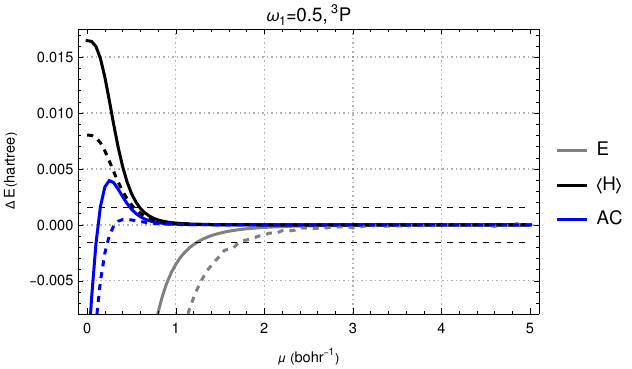}
    \caption{Total energy errors for the first excited state of harmonium, $^3P$, for $\omega=0.5$~a.u.; the description of the curves corresponds to that in fig.~\ref{fig:energies}.}
    \label{fig:excited}
\end{figure}
We consider the first excited state of harmonium.
It is a triplet state, dominated by an $sp$ configuration.
For triplet pairs, $\alpha(\mu)$ differs from that used for the singlet pair in the main body of the paper.
Electrons are better kept apart in the triplet state, and the corrections to the single Slater determinant are less important.
Although we see in Fig~\ref{fig:excited} at $\mu=0$ an improvement through the use of the OEP, its advantage becomes minimal when the error reaches chemical accuracy.

This open-shell case is different from that appearing in the case of near-degeneracy.
In this situation, the interaction $\op{W}(\mu)$ present in the Hamiltonian is supposed to take care of the coupling between the near-degenerate states.

\bibliography{biblio}
\end{document}